 \documentstyle[12pt]{article}
 
\begin{document}
\begin{flushright}
Preprint SU-GP-93/1-1\\
January 25, 1993
\end{flushright}

\begin{center}
{ {\bf  DIMENSION ON DISCRETE SPACES. }}
\end{center}
\bigskip
\begin{center}
{  Alexander V. Evako \\
 Syracuse University, Department of Physics \\
Syracuse
  N.Y. 13244-3901.
}
\end{center}
\bigskip

\begin{center}
 {\bf   Abstract}

\end{center}

In this paper we develop some combinatorial models for continuous
spaces. In this spirit we study the approximations of continuous
 spaces by graphs, molecular spaces and coordinate matrices.
We define the dimension on a discrete space by means of axioms, and
the axioms are based on an obvious geometrical  background.
This work presents some discrete models of n-dimensional   Euclidean
 spaces, n-dimensional spheres, a torus and a projective plane. It
explains how to construct new discrete spaces  and describes in this connection
 several three-dimensional closed surfaces with some topological singularities
 It also analyzes the topology  of (3+1)-spacetime.
We are also discussing the question by R. Sorkin [19] about how
 to derive the system of simplicial complexes from a system of open covering
of a topological space S.

 \subsubsection*{Introduction.}

A number of workers  have been unhappy about applications of the
continuum picture of space and spacetime. They have believed that the breakdown
of the functional integral  at the Plank length
shows not merely the failure of the classical field equations but also
indicates that a differential manifold upon which they are built should be
replaced by some finite theory. This was certainly one of the motivations
behind Penrose invention [15] of spin networks and recent works by Finkelstain
 on a novel spacetime microstructure [1].
Isham, Kubyshin and Renteln [2] introduce a quantum theory
 on the set of all
topologies  on a given set, and show that for a finite basic set almost all
metrics can be obtained by embedding this set into a vector space and
then varying the norm of this space.  \\
One more approach to a combinatorial model of space and spacetime is
studied in work [19] by R.Sorkin.
 He replaces general topological spaces by  a finite ones and describes how
to  associate a finite space with any locally finite covering of a
topological space.
He also presents some examples of posets derived from simple spaces.\\
Another way  is Regge calculus [18]. Many suggestions for formulating
various Regge calculus versions have been made in order to face a number of
 problems.  Regge calculus describes general
relativity spacetime
 by using a simplicial complex. Its fundamental variables are a set of
edge lengths and an
incidence matrix that describes how they are connected.
One approach supposes that the connectivity of a simplicial complex is
fixed, but the lengths of edges can be varied.
Another approach fixes the edge length and varies the connectivity of a
simplicial complex in order to change the metric of a spacetime.
The Regge calculus, however, supposes that there is a continuous
underlying spacetime and does not account naturally for the appearance
of a minimal
length in effective theories.\\
At the same time in mathematics there exist several quickly developed
approaches
 to discrete spaces in the frame of digital topology which can be useful in
physics. Digital topology is the study of topological properties of image
arrays.  It provides the theoretical foundations for image processing
operations such as image thinning, border following and object counting.\\
The paper [12] reviews the fundamental concepts of digital topology,
 surveys the major theoretical results in this field and contains the
bibliography of almost 140 references.\\
Traditionally  a  discrete or digital space  is considered as a graph whose
edges between vertices define the nearness and connectivity in the
neighborhood of any vertex.\\
This approach was used  by Rosenfeld [16,17] who proved the first version of
the Jordan curve theorem  by
using a graph theoretical model of a digital plane.
However, this model does not utilize a topological basis and requires different
nearness for the curve and its background.\\
An alternative topological approach to the digital topology uses the
notion of a connected topology on a totally ordered set $Z$ of integers
[9-11,13]. The digital plane $Z \times Z$ or the  three-dimensional digital
 space $Z \times Z \times Z$  are the topological products of two or three
such spaces respectively. Using this construction the Jordan
 curve theorem in two  and three dimensions  was proven.\\
Another approach to finite topology is offered by Kovalevsky in [14].
  He builds the digital space as a structure
consisting of elements of different dimensions by using a such  well-known in
topology element as a cellular complex.\\
Our approach to discrete spaces  is based on using three combinatorial tools:
a graph a molecular space
 and a coordinate matrix [3-8].
In this panel the material to be presented below begins with short
description of some  geometrical background for the definition of the
dimension on
a graph.
Then we shall show the connection between a graph, a molecular space and a
coordinate
 matrix.
We shall define the dimension on discrete spaces  which
is based on some geometrical ground.
We shall analyze the dimensions of different models
of two,  three and n-dimensional discrete  spaces.
We  present some
examples of three-dimensional discrete  closed spaces with strange topological
features which do not have direct continuous analogies.  Then
we prove some theorems showing how to construct closed three dimensional
spaces with nonstandard topology.  Finally we discuss
  the topological structure  of a (3+1) space-time.

 \subsubsection*{Geometrical background for the definition of the dimension on
a graph.}
We are going to construct now a graph with certain properties which can be
thought as a convenient tool for describing the ideas of nearness and
continuity by combinatorial methods. This will be done in the first
place by picking out in elementary geometry those properties of nearness
which seem to be fundamental and taking them as axioms.
To get a glimpse of the intuitive geometrical ground of the
dimension consider the following example.
Let $E^{n}$ be n-dimensional Euclidean space and $p$  a point in it.
The neighbourhood of $p$ is commonly defined to be any set $U$ such
that $U$ contains an open solid disk $D_{1}^{n}$ of center $p$.
The boundary of this disk is the sphere $S_{1}^{n-1}$.\\
The definition of neighborhood is formulated in this way so as
  to be as free as possible from the ideas of size and shape, concepts
that play no part in topology.\\
Using this definition  of a  neighbourhood of a $p$ in Euclidean space it
is easy to see that the family of sets $U$ satisfy the usual topological
axioms.\\
\begin{enumerate}
\item $p$ belongs to any  neighbourhood of $p$.
\item If $U$ is a  neighbourhood of $p$ and $U \subset V$, then $V$ is a
neighbourhood of $p$.
\item If $U$ and $V$ are  neighbourhood of $p$, so as is $U \bigcap V$.
\item If $U$ is a neighbourhood of $p$, then there is a neighbourhood
 $V$ of $p$ such that $V \subset U$ and $V$ is a
  neighbourhood of each
of its points.
\end{enumerate}
Taking these properties as axioms in an abstract formulations we can define
a topological space $E$ as a set $E$ with a family of subsets of $E$ satisfying
 the four properties listed earlier. We  can also define a subset $W$ of $E$
open if for each point $p$ in $W$, W is a neighbourhood of $p$.\\
Note that the disk $D_{1}^{n}$ plays crucial role in this definition.\\
In the continuous case the sphere $S_{1}^{n-1}$ contain in itself an infinite
sequence of disks $D_{i}^{n}$ and spheres $S_{i}^{n-1}$ of center $p$.
\[ D_{1}^{n} \supset D_{2}^{n} \supset....\supset D_{i}^{n} \supset....\]
\[ S_{1}^{n-1} \supset S_{2}^{n-1} \supset....\supset S_{i}^{n-1} \supset....\]
However the situation is different in the discrete case where the sequences
of disks and spheres can not be infinite and axiom 4 is not realized.
Therefore we have a finite series of the form
\[ D_{1}^{n} \supset D_{2}^{n} \supset....\supset D_{t}^{n}\]
\[ S_{1}^{n-1} \supset S_{2}^{n-1} \supset....\supset S_{t}^{n-1}\]
The smallest disk $D_{t}^{n}$ and the smallest sphere $S_{t}^{n-1}$ can not
be reduced in the sense that they do not contain disks and spheres  others
then themselves (Figure 1).\\
The topological meaning this construction for a graph reveals that the
vertex $p$ is considered as n-dimensional if its minimal neighbourhood is the
sphere $S^{n-1}$.\\
The point $p$ and the nearest sphere $S^{n-1}$ together form the smallest
disk $D^{n}$ of center $p$.
Point $p$ of a discrete space $G$ is considered as one-dimensional if its
nearest neighbourhood is a zero-dimensional sphere $S^{0}$. It is well
known that $S^{0}$ is a set of two disconnected points. In the other
words $S^{0}$ is disjoined graph of two points.
In one-dimensional discrete sphere $S^{1}$ all points are one-dimensional.
Obviously the minimal number of points required for $S^{1}$ is four.
 For two-dimensional discrete sphere $S^{2}$ all its points are
two-dimensional. It means that nearest  neighbourhood of any point of $S^{2}$
should be $S^{1}$ and so on.

\subsubsection*{ A molecular space  and a coordinate matrix of
 a graph. }

In order to make this paper self-contained we shall summarize the necessary
results from our previous papers.
Let $E^\infty $ be infinite-dimensional Euclidean space. Take the
coordinates of a point $x,\;\;  x\in E^\infty $, as a sequence of real numbers
\[x=(x_{1},x_{2},..,x_{n},..)=[x_{i}],\;\;  i\in N.  \]
We define unit cube $K\in E^\infty $ in the following way: each $x,\;\;
x\in K$,
has coordinates ${x_{i}}$ satisfying conditions presented in [3,4,5,8]:

\[   n_{i}\leq x_{i}\leq n_{i}+1,\;\; i\in N,\;\; n_{i} - integer.\]

Therefore, $K$ is an infinite-dimensional cube with unit edges. In
[3,4,5,8] $K$ is called  a kirpich. We will use this name in the
present paper. \\
 The position of $K$ in  $E^\infty $ is determined by the left
vertex coordinates.  For the given kirpich we have

  \[   K=(n_{1},n_{2},...,n_{n},...)=[n_{i}],\;\; i\in N.\]

Two kirpiches are called adjacent, if they have common points.  The
distance $d(K_{1},K_{2})$ between kirpiches $K_{1}=[n_{i}]$ and $K_{2}=[m_{i}]$
is defined by using sup norm

		\[d(K_{1},K_{2})=max|n_{i}-m_{i}|,\;\; i\in N.\]

Obviously, two kirpiches are adjacent if their appropriate coordinates
distinguish not more then 1, or the distance between them equals 1. Any
set of kirpiches in $E^\infty$ is  called a molecular space and is denoted by
 $M$. Clearly, any molecular space can be represented by its
intersection graph $G(M)$. It was shown in [4,8] that any graph $G$ can be
represented by a molecular space $M(G)$, such that $G=G(M(G))$. Clearly,
more than one $M(G)$ can be built for the graph $G$. There exists
isomorphism between any two  $M(G)$. \\
 Let  $M$
 be a molecular space  with a set of kirpiches
 \[ V=(K_{1},K_{2},..,K_{n}), \;\;\; K_{1}=[k_{1i}],
\;\;K_{2}=[k_{2i}],...K_{n}=[k_{ni}]  \]
The matrix $[k_{pi}] \;\;$  is called the coordinate matrix of the  molecular
  space $M$  and  its intersection graph
$G(M)$  and  is denoted $A(M)$ or $A(G(M))$. This matrix has $n$ rows
and infinite columns.\\
In fact we shall always use a finite-dimensional Euclidean space.
The intuitive background for using the infinite-dimensional unit cube
is the attempt to create some universal element not depending on the
dimension and suitable for describing elements of different dimensions:
zero-dimensional points, one-dimensional lines, two-dimensional surfaces
and  so on.\\
Let $S$ be a surface in $E^{n}$. The molecular space $M(S)$ of $S$ is
a set of kirpiches intersecting $S$.\\
Figure 2 shows the graph $G$, its molecular space $M$ and its coordinate
matrix $A$.

\subsubsection*{The dimension and the metric of a discrete space .}
Our objective now is to define the dimension on graphs.
 Later on we will use the names discrete space and point for a graph
and its vertex
when we want to emphasize the notion of the dimension on it.\\
Since in this paper we only use induced subgraphs, we shall use the word
subgraph for an  induced subgraph. We shall also use some
symbols, notations and names introduced in our previous works.

 \subparagraph*{Definitions}
  Let  $G,$ $G_{1}$ and $v$
  be a graph, its subgraph and its point.\\
$\bullet$ The subgraph $B(G_{1})$  containing  $G_{1}$ is called the ball of
$G_{1}$ if any  point of $B(G_{1})$ is adjacent to at least one point
of $G_{1}$.\\
$\bullet$ The subgraph $B(G_{1})$ without points of $G_{1}$ is called the
rim of $G_{1}$ and it is denoted $O(G_{1}).$\\
Obviously $B(G_{1})-G_{1}=O(G_{1}).$\\
$\bullet$ If $G_{1}$ is a point $v$ then $B(v)$ and $O(v)$ are called the
ball and the rim of $v$ respectively.\\
$\bullet$ The subgraph $B(v_{1},v_{2},...v_{n}),$  $B(v_{1},v_{2},...v_{n})=
B(v_{1}) \bigcap B(v_{2}) \bigcap... B(v_{n}),$  is called the joint
 ball of the points $v_{1},v_{2},...v_{n}$.\\
$\bullet$ The subgraph $O(v_{1},v_{2},...v_{n}),$  $O(v_{1},v_{2},...v_{n})=
O(v_{1}) \bigcap O(v_{2}) \bigcap... O(v_{n}),$  is called the joint
 rim of the points $v_{1},v_{2},...v_{n}$.\\
Let $G$ and $G_{1}$ be a graph and its subgraph with points
$(v_{1},v_{2},...v_{n})$ and  $(v_{1},v_{2},...v_{p})$  respectively.
It is clear that
\[ O(G_{1})=O(v_{1}) \bigcup O(v_{2})
\bigcup...\bigcup O(v_{p})-G_{1}\]
\[   B(G_{1})=B(v_{1}) \bigcup B(v_{2})
\bigcup...\bigcup B(v_{p}) \]
$\bullet$ A  graph $K_{n}(v_{1},v_{2},...v_{n})$ of $n$
points is called completely connected or complete if any its two points are
adjacent.\\
$\bullet$  A  graph $H_{n}(v_{1},v_{2},...v_{n})$ of $n$
points is called completely disconnected  if any its two points are
disjoined.\\
$\bullet$ The join $G_{1}*G_{2}$ of two graphs $G_{1}$ and $G_{2}$
is the graph which consists of two graphs $G_{1}$ and $G_{2}$ and all edges
joining points of $G_{1}$  with points of  $G_{2}$.\\
To this end we begin by defining the dimension on a graph
in the following way.
\begin{description}
\item[Definition 1] Zero-dimensional normal space $S^{0}$ is the graph which
consists
of two non-adjacent  points.\\

\item[Definition 2] A point $v$ of a graph $G$ is called a normal n-dimensional
 point if its rim $O(v)$ is a normal (n-1)-dimensional space.\\

\item[Definition 3] For any integer $n$, $n \geq 1$, define a normal
n-dimensional
space to be a connected graph in which any  point is n-dimensional normal.

\end{description}

 According to these definitions one-dimensional normal space is any circle
 $C_{n},\;\; n \geq 4.$ \\
Further the denotation $p(G)$ will be used
for the dimension of a  graph $G$.\\
Figure 3  represents normal zero, one and two-dimensional spheres and
   one, two and three-dimensional disks.\\
Figure 4 shows normal  two-dimensional discrete flat spaces and their
molecular spaces. $E_{1}^{2}$ is the two-dimensional discrete space in
Khalimsky topology [9-13].\\
Tree-dimensional normal sphere is the graph $S^{3}$ depicted in Figure 5. It
can be
verified without difficulties that the complete (n+1) partite graph
$K(2,2,...2)$ is the minimal graph describing $S^{n}$ [3,4,6]. Therefore, the
minimal number of elements necessary to describe $S^{n}$ is 2n+2. Notice that
the same number of points is used by R. Sorkin [19] to describe $S^{n}$ in
the finitary topology approach.\\
A normal  torus $T^{2}$  and  a projective plane $P^{2}$ are presented
in Figure 5.\\
It can be checked directly that the Euler characteristic  and the homology
groups  of all graphs  depicted in Figures 3-5  match the Euler
characteristic and the homology groups of their continuous
counterparts [5,7].\\
In [3] normal n-dimensional space is called of the type $\Pi_{2}$. This
separation to the different types is caused by the fact the normal
molecular spaces and graphs of any type $\Pi_{n},\;\;n \neq 1,2$
have some unusual properties different from those of
direct discrete models
of continuous spaces in $E^{m}$.

Our objective now is to define a generalization of the dimension
which includes the above definition. It is natural to consider a
point $v$ as zero-dimensional if its neighborhood does not contain any
normal space.
\begin{description}
\item[Definition 4] A point $v$ of a graph $G$ is called zero-dimensional,
 $p(v)=0$, if $O(v)$ does not contain
the normal  zero-dimensional sphere $S^{0}$.
\item[Definition 5] A connected graph $G$ is called zero-dimensional,
$p(G)=0$, if any of its points is zero-dimensional.
\end{description}
By this definition in a zero-dimensional connected graph any two points are
adjacent. Therefore, this graph is a complete graph on any number of
vertices.  A disconnected zero-dimensional   graph is considered
as a zero-dimensional sphere $S^{0}$ if it has exactly two components.
It is clear that $S^{0}$ contains normal zero-dimensional sphere as its
subgraph. We will extend this analogy to higher dimensions.
\begin{description}
\item[Definition 6] A graph $G$  is called closed n-dimensional if\\
1. For any point $v\;\;\;p(v) \leq n$.\\
2. $G$  is homotopic to some  normal n-dimensional space.\\
\item[Definition 7] A point $v$ is called n-dimensional, $p(v)=n$, if\\
1.  $O(v)$ contains  a closed (n-1)-dimensional space.\\
2.   $O(v)$ does not contain any  closed   n or more-dimensional space.\\
\item[Definition 8] A graph $G$  is called n-dimensional, $p(G)=n$, if\\
1. $G$  contains at least one n-dimensional point\\
2. For any point $v\;\;\;p(v) \leq n$.\\
\end{description}
In definition 6 we use homotopy of graphs. Two graphs are called homotopic
if each of them can be turned into the other by contractible transformations
  which consist of contractible gluing and deleting of vertices and
edges of a graph. It was shown [5-7] that these transformations do
not change the Euler characteristic and the homology groups of graphs.\\
Let us look at some examples of n-dimensional (not normal) discrete
spaces  and their
molecular
spaces.\\
Spheres $S^{0}$, $S^{1}$,  their molecular spaces and the molecular space
$M(S^{2})$ of sphere $S^{2}$
are drawn in Figure 6.  $M(S^{2})$ is a hollow space, it does not
contain the central unit cube.
These spheres are not normal but satisfy definitions
6-7. Any sphere $S^{n}$ depicted in Figure 6 has the same Euler characteristic
and homology groups as continuous $S^{n}$ and can be transformed to the
sphere $S^{n}$ drawn in Figure 3  by contractible transformations.
Flat one, two and three-dimensional  spaces and their molecular
spaces are
shown in Figure 7. It is easy to construct three and more dimensional
 spaces but it is difficult to draw it. For a flat
three-dimensional  space  the only molecular space  is shown.
However a  n-dimensional space can be easily  described by its
coordinate matrix of the form
\[\left(
\begin{array}{cccc}
x_{11} & x_{12}& \cdots & x_{1n}\\
x_{21} & x_{22}& \cdots & x_{2n}\\
\vdots & \vdots & \ddots & \vdots \\
x_{p1} & x_{p2}& \cdots & x_{pn}\\
\cdots & \cdots & \cdots & \cdots
\end{array}
\right )
\]
where $x_{ik}=0,\pm 1, \pm 2,...; \;\;\;i=1,2,3,..; \;\;\;k=1,2,...n.$ \\
The standard definition of the distance on a graph can be applied to a
discrete space.
\begin{description}
\item[Definition 9] The distance $d(v_{1},v_{2})$ between two points
 $v_{1}$ and $v_{2}$ in a discrete space $G$ is the length of a shortest
path joining
 them if any; otherwise $d(v_{1},v_{2})= \infty$.
\end{description}
Obviously the distance is a metric. The Plank length can be thought as the
length of an edge of the  graph.

 \subsubsection*{Mathematical observations.}
Before proceeding to the main result of this paper let us pause to describe
some mathematical observations relating to this approach.\\
The following surprising facts were revealed.\\
$\bullet$ Suppose that $S^{1}$ is a circle of radius $R$. Let $A$ be a cover of
 $S^{1}$ by arcs whose length is small enough compared  with $R$.
Denote $G(A)$ the intersection graph of this cover. This graph is called the
circular arc graph. It appears that :\\
1. Dimension of  $G(A)$ is equal to one, $p(G(A))=dim(S^{1})=1$.\\
2. $G(A)$ has the same Euler characteristic and homology groups as
$S^{1}$.\\
3. $G(A)$ can be reduced to the cycle graph $C_{4}$  by
contractible transformations [5,6,7] ($S^{1}_{1}$ in Figure 3).\\
$\bullet$ Suppose we have some two-dimensional closed surface, for example,
a sphere $S^{2}$ of radius $R$.  Consider any tiling $A$ of $S^{2}$ by elements
$(a_{1},a_{2},...a_{n})$ whose size is small enough relative to the radius $R$.
 Construct the intersection graph $G(A)(v_{1},v_{2},...v_{n})$ in the following
 way: Two vertices $v_{1}$ and $v_{2}$ are adjacent iff elements $a_{1}$ and
$a_{2}$ have at least one common point. In most cases it turns out that \\
1. Dimension of  $G(A)$ is equal to two, $p(G(A))=dim(S^{2})=2$.\\
2. $G(A)$ has the same Euler characteristic and homology groups as
$S^{2}$.\\
3. $G(A)$ can be reduced by contractible transformations into the
minimal two-dimensional sphere on 6 vertices [5,6,7] ($S^{2}$ in Figure 3).\\
$\bullet$ Suppose that $P^{k}$ is a  surface in $E^{n},\;\;n=2,3$ (for
spheres $n$
 can be any number).
 Divide $E^{n}$ into a set of  cubes with the scale $l_{1}$
 of the cube edge and call the molecular space $M_{1}(P^{k})$ of $P^{k}$
  the family of cubes intersecting $P^{k}$. Denote $G_{1}(P^{k})$ the
intersection graph of $M_{1}(P^{k})$.
                Change the scale of the cube edge from   $l_{1}$  to
  $l_{2}$  and obtain  $M_{2}(P^{k})$  and  $G_{2}(P^{k})$
by using
the same structure.
 It is revealed that in most of cases\\
1. $p(G_{1}(P^{k}))=p(G_{2}(P^{k}))=dim(P^{k})$\\
2. $G_{1}(P^{k})$ and $G_{2}(P^{k})$ have the same Euler characteristic and
the homology groups as $P^{k}$.\\
3.  $G_{1}(P^{k})$ and $G_{2}(P^{k})$ can be transformed
from one to the other with four kinds of transformations if the  divisions are
small
 enough.\\
$\bullet$$\bullet$These facts allow us to assume that the graph and the
molecular space  contain topological and perhaps geometrical characteristics of
the surface $P^{k}$.  Otherwise, the molecular space $M$ and the graph $G$
are  the discrete counterparts of a continuous space $P^{k}$.

 \subsubsection*{Singular spaces.}
This section describes a method of obtaining new spaces from given ones.
We will see that there exist n-dimensional normal spaces with some
 peculiar properties. These spaces give rise to new discrete structures
that have different topologies in different points.
\begin{description}
\item[Theorem 1] Let  $G(p_{1},p_{2},...p_{r})$  and  $H_{2}(v_{1},v_{2})$ be
a n-dimensional normal space and the
completely disconnected space on two points respectively. Then
$H_{2}(v_{1},v_{2})*G(p_{1},p_{2},...p_{r})$ is a
(n+1)-dimensional normal space.
\item[Proof.] The proof is by induction. \\
(i) For n=0,1 the theorem is verified directly. \\
(ii) Assume that theorem is valid for any n, $n \leq k$. Let
 $G(p_{1},p_{2},...p_{r})$ be a normal (k+1)-dimensional discrete space.
Consider
\[  W=H_{2}(v_{1},v_{2})*G(p_{1},p_{2},...p_{r}) \] It is necessary
to show that $W$ is a (k+2)-dimensional discrete
normal space. Take any point $p_{i}$. With respect to the
definition of a
normal space, $O(p_{i})$ in $G$ denoted
$O(p_{i})|G$ is a k-dimensional normal
space. Therefore, according to the assumption $H_{2}*O(p_{i})$
is (k+1)-dimensional
normal space. Hence any point $p_{i}$ in $W$ has the rim
 which is (k+1)-dimensional normal space. \\
The rims of points $v_{1}$ and $v_{2}$ in $W$ are the (k+1)-dimensional
normal space $G$
 by construction.
\[ O(p_{i})|W=H_{2}*(O(p_{i})|G),\;\;i=1,2,...n,\;\;\;O(v_{k})|W=G,\;\;k=1,2
\]
Therefore, the rim of any point of $W$ is a (k+1)-dimensional
 normal  space, and, by the definition, $W$ is a normal (k+2)-dimensional
space. That completes the proof. $\Box$\\
\end{description}
We are now in a position to describe n-dimensional normal spaces with
peculiar properties. \\
$\bullet$ Firstly construct a space without singularities. Let $G$ be
n-dimensional sphere $S^{n}$. It means that the rim of
any  point of $S^{n}$ is a normal sphere $S^{n-1}$, and $S^{n}$ can be turned
into the minimal $S^{n}$ on 2n+2 points by contractible transformations
[3,6,7].\\
Consider $W$=$H_{2}(v_{1},v_{2})*S^{n}$. If $p \in S^{n}$ then
$O(p)|W=H_{2}(v_{1},v_{2})*S_{p}^{n-1}=S_{p}^{n}$. For points $v_{1}$ and
$v_{2}$ the rim is $S^{n}$ itself. Therefore, the rim of any point of
$W$ is sphere   $S^{n}$, and $W$ is a normal (n+1)-dimensional space.
It is easy to show   that $W$ can be reduced to the minimal (n+1)-sphere
$S^{n+1}$ by contractible transformations and, therefore, $W=S^{n+1}$.\\
 \[ W=H_{2}*S^{n}=S^{n+1} \]
\[ O(p)|W=H_{2}*S_{p}^{n-1}=S_{p}^{n},\;\;\;\;p \in S^{n}; \;\;\;\;\;\;\;
O(v_{1})|W=O(v_{2})|W=S^{n} \]
             \
$\bullet$ Suppose that $G$ is a discrete two-dimensional torus $T^{2}$ depicted
in Figure 5. For any point  $p$ of $T^{2}$ $O(p)=S_{p}^{1}$. Therefore, in
$W=H_{2}(v_{1},v_{2})*T^{2}$ the rim of any point $p$ is a two-dimensional
sphere
$S_{p}^{2}$, $O(p)|W=S_{p}^{2}$.  However, for points $v_{1}$ and $v_{2}$
their rims are the torus
$T^{2}$ itself, $O(v_{i})=T^{2}$, i=1,2. Notice that the dimension of $T^{2}$
is equal to 2.
 Hence $W$ is a normal three-dimensional space in which the rims of points have
 a different  topology. For points $v_{1}$ and $v_{2}$ the space has torus
neighborhood $T^{2}$, in all other points the neighborhood is spherical,
$S^{2}$.
\[  W=H_{2}*T^{2} \]
\[ O(p)|W=H_{2}*S_{p}^{1}=S_{p}^{2},\;\;\;\; p \in T^{2};\;\;\;\;\;\;
O(v_{1})|W=O(v_{2})|W=T^{2} \]

$\bullet$ Another peculiar three-dimensional space appears when we choose
the projective plane $P^{2}$ (Figure 5) as a basic space $G$.\\
In three-dimensional normal space $W=H_{2}(v_{1},v_{2})*P^{2}$ the
neighbourhoods
of $v_{1}$ and $v_{2}$ are the projective plane $P^{2}$, the neighbourhoods of
all other points are usual spheres $S^{2}$.\\
\[ W=H_{2}*P^{2} \]
\[ O(p)|W=H_{2}*S_{p}^{1}=S_{p}^{2},\;\;\;\;p \in P^{2}; \;\;\;\;\;\;
O(v_{1})|W=O(v_{2})|W=P^{2} \]
          $\bullet$ In general we can create a number of three-dimensional
normal spaces with two
singularities by taking discrete models of closed  two-dimensional oriented or
non-oriented surfaces as a basic space.

 \subsubsection*{The dimensional local structure of a physical discrete
(3+1) space-time.}

Now we are ready to discuss some general features of the physical
 (3+1) space-time.
We will restrict our consideration by local properties of a point $v$.
\begin{description}
\item[Theorem 2.] (3+1) space-time is four-dimensional non-normal.
\item[Proof.] We have to prove that in (3+1) space-time the rim of any point
is a closed three-dimensional  non-normal  discrete space. \\
Suppose that a physical object is in point $v$ of a three-dimensional discrete
space $R(t)$ at  a given moment $t$ and at either the same or the nearest point
 $v_{1}$ at the next moment $t+Dt.$ (Figure 8a). In (3+1)  space-time
$(R,T)$ we have two three-dimensional spaces
$R(t)$ and
$R(t+Dt)$  corresponding to the different moments. Obviously these
 spaces are joined together in the following way. Point $v$ on $R(t)$ should
be connected with the ball $B(v)$ on $R(t+Dt)$ (Figure 8b).
Therefore, in the (3+1)  space-time $(R,T)$ (Figure 8c)
the rim  $O(v)|(R,T)$  of
point $v$ is as shown in Figure  8d.\\
(i) If  the rim  $O(v)|R(t)$  of $v$  in $R(t)$  is a non-normal closed
two-dimensional
space, then, for the same reasons as in theorem 1,  $O(v)$ in $(R,T)$ is a
non-normal closed three-dimensional space,
 and $(R,T)$
is a non-normal four-dimensional space.\\
(ii) Suppose that $R(t)$ is a normal three-dimensional space.
Then $O(v)$ in $R(t)$ is a normal two-dimensional discrete space.
Obviously $O(v)$ in $(R,T)$ contains the normal three-dimensional space
$H(u_{1},u_{2})*O(v)|R(t)$ where $u_{1}$ and $u_{2}$ are $v$ in $R(t+Dt)$
and $R(t-Dt)$. By theorem 1 it is a normal three-dimensional space.
Take $v_{1}$ in $R(t+Dt)$, $v_{1} \in R(t+Dt),\;\;v_{1} \in O(v)|(R,T)$.
It is easy to see that   $v$ in $R(t+Dt)$
 is adjacent to all
points of the rim of this $v_{1}$ in $O(v)|(R,T)$.
Hence $O(p)|(R,T)$ is a non-normal closed three-dimensional space which
can be reduced into normal $H(u_{1},u_{2})*O(v)|R(t)$  by contractible
 transformations. Thus $(R,T)$ is a non-normal four-dimensional space-time.
 $\Box$\\
\end{description}

\subsubsection*{Acknowledgment}
The author would like to thank Rafael Sorkin  for his
useful  comments which resulted in several improvements
of the presentation of this paper.

\subsubsection*{ References}

\begin{enumerate}

\item D. Finkelstain, First flash and second vacuum,  International
Journal of Theoretical Physics 28 (1989) 1081-1098.

\item C. J. Isham, Y. Kubyshin and P. Renteln, Quantum norm theory
and the quantisation of metric topology,  Classical and
Quantum Gravity 7 (1990) 1053-1074.

\item A.V. Ivashchenko (Evako), Dimension of molecular spaces, VINITI, Moscow,
    6422-84 (1985) 3-11.

\item  A.V. Ivashchenko (Evako), Geometrical representation of a graph,
Kibernetika
    5 (1988) 120-121.

\item A.V. Ivashchenko (Evako), Homology groups on molecular spaces
and graphs, Discrete Math., to appear.

\item A.V. Ivashchenko (Evako), Yeong-Nan Yeh, Minimal graphs of a torus,
a projective plane and spheres. Some properties of minimal graphs of
homology classes,  Discrete Math., to appear.

\item A.V. Ivashchenko (Evako), Representation of smooth surfaces by graphs.
Transformations of graphs which do not change the Euler characteristic
of graphs, Discrete Math., 122 (1993) 219-233.

\item A.V. Ivashchenko (Evako), The coordinate representation of a graph and
    n-universal graph of radius 1, Discrete Math., 120 (1993) 107-114.

\item E. Halimski (E. Khalimsky), Uporiadochenie Topologicheskie
     Prostranstva (Ordered Topological Spaces), Naukova Dumka, Kiev, 1977.

\item E. Khalimsky, R. Kopperman, P. Meyer, Computer graphics and
                   connected topologies on finite ordered sets, Topology and
    Applications 36 (1990) 1-17.

\item T. Kong, P. Meyer, R. Kopperman, A topological approach to digital
topology, Amer. Math. Monthly 98 (1991) 901-917.

\item T. Kong, A. Rosenfeld, Digital topology: Introduction and survey,
    Comput. Vision Graphics Image Process 48 (1989) 357-393.

\item R. Kopperman, P. Meyer, R. Wilson, A Jordan surface of three-
    dimensional digital space, Discrete Comput Geom. 6 (1991) 155-161.

\item V.A. Kovalevsky, Finite topology as applied to image analyses,
    Comput. Vision Graphics Image Process 46 (1989) 141-161.

\item R. Penrose,  Angular momentum: an approach to combinatorial space-time.
 in "Quantum theory and beyond", ed. by T. Bastin ( Cambridge University
 Press, Cambridge, 1971).

\item A. Rosenfeld, Connectivity in digital pictures, J. Assoc. Comput.
 Mach. 17 (1970) 146-160.

\item A. Rosenfeld, Digital topology, Amer. Math.
Monthly 86 (1979) 621-630.

\item T. Regge, General relativity without coordinates, Nuovo
Cimento 19 (1961) 558-571.

\item R. Sorkin, Finitary substitute for continuous topology, International
Journal of Theoretical Physics 30 (1991) 923-947.

\end{enumerate}

 \subsubsection*{Figure Captions.}

Figure 1:  Difference between an infinite and finite number of enclosed disks
$D^{n}$ in continuous and discrete spaces respectively.\\

Figure 2: Graph $G$, its molecular space $M(G)$ and its coordinate matrix
$A(G)$.\\

Figure 3: Zero $(S^{0})$, one  $(S^{1}_{1}, S^{1}_{2})$ and two $(S^{2})$
 dimensional
 normal discrete spheres, and one $(v_{1})$, two
$(v_{2}, v_{3})$ and three $(v_{4})$ dimensional points.\\

Figure 4: Normal discrete two-dimensional planes and their molecular spaces.
$E^{2}_{1}$ is the two-dimensional plane in Khalimsky topology.\\

Figure 5: A discrete normal three-dimensional sphere $S^{3}$, a two-dimensional
torus $T^{2}$, a  two-dimensional projective plane $P^{2}$. The Euler
characteristic and the homology groups of these graphs are consistent
with the Euler
characteristic and the homology groups of their continuous counterparts.\\

Figure 6: Zero and one-dimensional non-normal spheres $S^{0}$ and $S^{1}$
and their molecular spaces $M(S^{0})$ and    $M(S^{1})$.
$M(S^{2})$ is a molecular space of the two-dimensional non-normal
sphere $S^{2}$.
It does not contain the central unit cube.\\

Figure 7: Non-normal  discrete one and two-dimensional flat spaces $E^{1}$ and
$E^{2}$  and their molecular spaces $M(E^{1})$  and $M(E^{2})$. $M(E^{3})$
is a molecular space of a non-normal  discrete three-dimensional flat
space  $E^{3}$.\\

Figure 8: Theorem 2 for (1+1) space-time.
  (1+1) space-time is not normal because $O(v)$ is not a normal
one-dimensional sphere.

\end{document}